\documentclass[lettersize,journal]{IEEEtran}
\usepackage{graphicx}
\usepackage{amsmath,amsfonts}
\usepackage{algorithmic}
\usepackage{algorithm}
\usepackage{array}
\usepackage[caption=false,font=normalsize,labelfont=sf,textfont=sf]{subfig}
\usepackage{amssymb}
\usepackage{textcomp}
\usepackage{stfloats}
\usepackage{url}
\usepackage{multirow}
\usepackage{verbatim}
\usepackage{afterpage} % 确保表格与内容不重叠
\usepackage{graphicx}
\usepackage{upgreek}
\usepackage[%
  top=0.75in,%
  bottom=1in,%
  left=0.625in,%
  right=0.625in%
]{geometry}
\usepackage{xcolor}
\usepackage{cite}
\begin{document}

\makeatletter
\newcommand{\rmnum}[1]{\romannumeral #1}
\newcommand{\Rmnum}[1]{\expandafter\@slowromancap\romannumeral #1@}
\makeatother

\title{Integrated Sensing and Communication \\ for 6G Holographic Digital Twins}

\author{Haijun Zhang,~\IEEEmembership{Fellow,~IEEE,}
        % <-this % stops a space
        \and
        Ziyang Zhang,
        \and
        Xiangnan Liu,~\IEEEmembership{Member,~IEEE,}
        \\ \and
        Wei Li,
        \and
        Haojin Li,
        \and
        and Chen Sun,~\IEEEmembership{Senior Member,~IEEE}

\thanks{ Haijun Zhang, Ziyang Zhang, and Wei Li are with Beijing Engineering and Technology Research Center for Convergence Networks and Ubiquitous Services, University of Science and Technology Beijing, Beijing, China, 100083. (email: zhanghaijun@ustb.edu.cn, m202310638@xs.ustb.edu.cn, li\_wei@ustb.edu.cn).}
\thanks{ Xiangnan Liu is with the School of Electrical Engineering and Computer Science, KTH Royal Institute of Technology, Stockholm, Sweden, 11428. (email:  xiangliu@kth.se).}
 \thanks{
 Haojin Li and Chen Sun are with Sony China Research Laboratory, Beijing,
 China, 100027. (email: Haojin.li@sony.com, Chen.Sun@sony.com).}}

\maketitle

\begin{abstract}
With the advent of 6G networks, offering ultra-high bandwidth and ultra-low latency, coupled with the enhancement of terminal device resolutions, holographic communication is gradually becoming a reality. Holographic digital twin (HDT) is considered one of key applications of holographic communication, capable of creating virtual replicas for real-time mapping and prediction of physical entity states, and performing three-dimensional reproduction of spatial information. In this context, integrated sensing and communication (ISAC) is expected to be a crucial pathway for providing data sources to HDT. This paper proposes a four-layer architecture assisted by ISAC for HDT, integrating emerging paradigms and key technologies to achieve low-cost, high-precision environmental data collection for constructing HDT. Specifically, to enhance sensing resolution, we explore super-resolution techniques from the perspectives of parameter estimation and point cloud construction. Additionally, we focus on multi-point collaborative sensing for constructing HDT, and provide a comprehensive review of four key techniques: node selection, multi-band collaboration, cooperative beamforming, and data fusion. Finally, we highlight several interesting research directions to guide and inspire future work.
\end{abstract}

\section{Introduction}
\textcolor{black}
{
The sixth-generation mobile communication system (6G) will realize the intelligent interconnection of humans, machines, and objects. The 6G wireless system is anticipated to support a peak data rate of 1 Tbps, ultra-high reliability with a packet error rate of \(10^{-9}\), and an end-to-end latency of 0.1 milliseconds, possessing the capability to stream holographic data from remote sites. Concurrently, as high-resolution three-dimensional display devices continue to evolve, holographic communication is moving toward reality \cite{1}. Holographic communication, through real-time collection, transmission, and rendering, reconstructs three-dimensional information, establishing holographic image connections for entities distant in the physical world through data transmission channels, thereby defining new modes of communication and interaction paradigms \cite{2}.
}

\textcolor{black} {As one of the significant applications of holographic communication, Digital Twin (DT) can greatly enhance the practicality and experience of holographic communication by integrating and virtually mapping real-world data. The usability of DT is largely influenced by its representation form. Although the capture, rendering, and display technologies of two-dimensional images and videos have matured, their inability to display depth information fails to provide a true experience for human visual perception. Consequently, Holographic Type Communication (HTC), leveraging three primary technologies: XR head-mounted display, multi-view volumetric display, and lightfield display, to interpret multi-perspective 3D holograms, will be closely integrated with DT in the future, giving rise to Holographic Digital Twins (HDT) \cite{6}. Unlike traditional DT, which are primarily used for monitoring, predicting, and optimizing the performance of physical entities, with user interaction limited to data analysis and the viewing of simulation results, HDT will allow users to immerse themselves in a twin world derived from the real world from various angles, while approaching the experience in the physical world with high resolution and high frame rates.}

\textcolor{black} {The realization of HDT relies on a comprehensive understanding and real-time monitoring of their physical counterparts, a foundation built upon the acquisition of precise environmental sensing data. Utilizing cameras to collect data on objects and their surroundings necessitates the deployment of numerous expensive devices and suffers significantly under low-light conditions. Integrated Sensing and Communication (ISAC) has recently emerged as a candidate technology for 6G wireless networks, enabling Sensing and Communication (S\&C) to share wireless infrastructure and spectrum resources. This provides high-rate communication and high-resolution sensing at an exceedingly low cost, capable of perceiving and reconstructing the environment around unmanned aerial platforms and wireless access points without being affected by weather and lighting conditions \cite{8}. If further supplemented with cameras, GPS, and other sensors for multimodal fusion, it could achieve complementary modalities, such as capturing color visual information with cameras while ISAC provides high-precision distance and speed measurements. This approach enables the acquisition of richer environmental information, thereby constructing a more realistic and comprehensive twin world and achieving a leap in HDT performance \cite{3}.}

\textcolor{black} {The construction of three-dimensional environments constitutes a crucial facet of HDT, necessitating the amalgamation of data from multiple sensing nodes positioned at various angles to restore a comprehensive 3D environment, as a single sensing node can only perceive a fragment of the environmental information \cite{4,8}. Moreover, the presence of obstructions in the environment often results in Non-Line of Sight (NLOS) conditions between sensing nodes and their targets, preventing direct signal propagation to the intended targets. Against this backdrop, there is an urgent need to explore various aspects of collaborative ISAC to achieve more extensive perception and enhance the accuracy of sensing, thereby providing adequate technological support for ISAC-assisted HDT 3D environment construction. In this article, we will delve further into specific aspects of multi-cooperative sensing, laying a more comprehensive theoretical foundation for multi-node collaborative sensing employed in the construction of HDT.}

\textcolor{black} {The rest of this paper is organized as follows: We start by discussing the use cases and KPIs of HDT supported by ISAC. We then propose a four-layer architecture for ISAC-assisted HDT and introduce super-resolution techniques in the physical layer and the intelligence layer to enhance sensing resolution.} Our focus is on collaborative sensing for HDT construction, including simulations on the relationship between node count, S\&C performance, and power needs. Finally, we conclude with a brief overview of potential challenges and future directions for constructing HDT through ISAC data acquisition.

\section{ISAC-Assisted Holographic \\ Digital Twin Solution}
In this section, we introduce the application scenarios of HDT firstly. And then the general framework of HDT is elaborated, assisted by ISAC.

\subsection{Overview of ISAC-HDT}
In the smart city scenario depicted in Fig. 1, the city is outfitted with Sub-6GHz and millimeter-wave base station (BS), working alongside unmanned aerial vehicles (UAV), which act as aerial BS, to establish an ISAC network. \textcolor{black}{Inside buildings, terahertz wireless access points are utilized.} This setup employs  \textcolor{black} {cooperative multi-point joint transmission (CoMP-JT)} for communication and multi-node cooperative sensing for data collection. The aggregated sensing data from these various sources are then utilized to create DT, facilitating immersive applications via holographic video. UAV, with their computational tasks related to sensing data, may offload heavier processing to edge servers due to their limited computational capabilities. These edge servers are crucial for merging sensing data from multiple points and for the computation and crafting of DT models.

\begin{figure}[!t]
\centering
\includegraphics[width=1\linewidth]{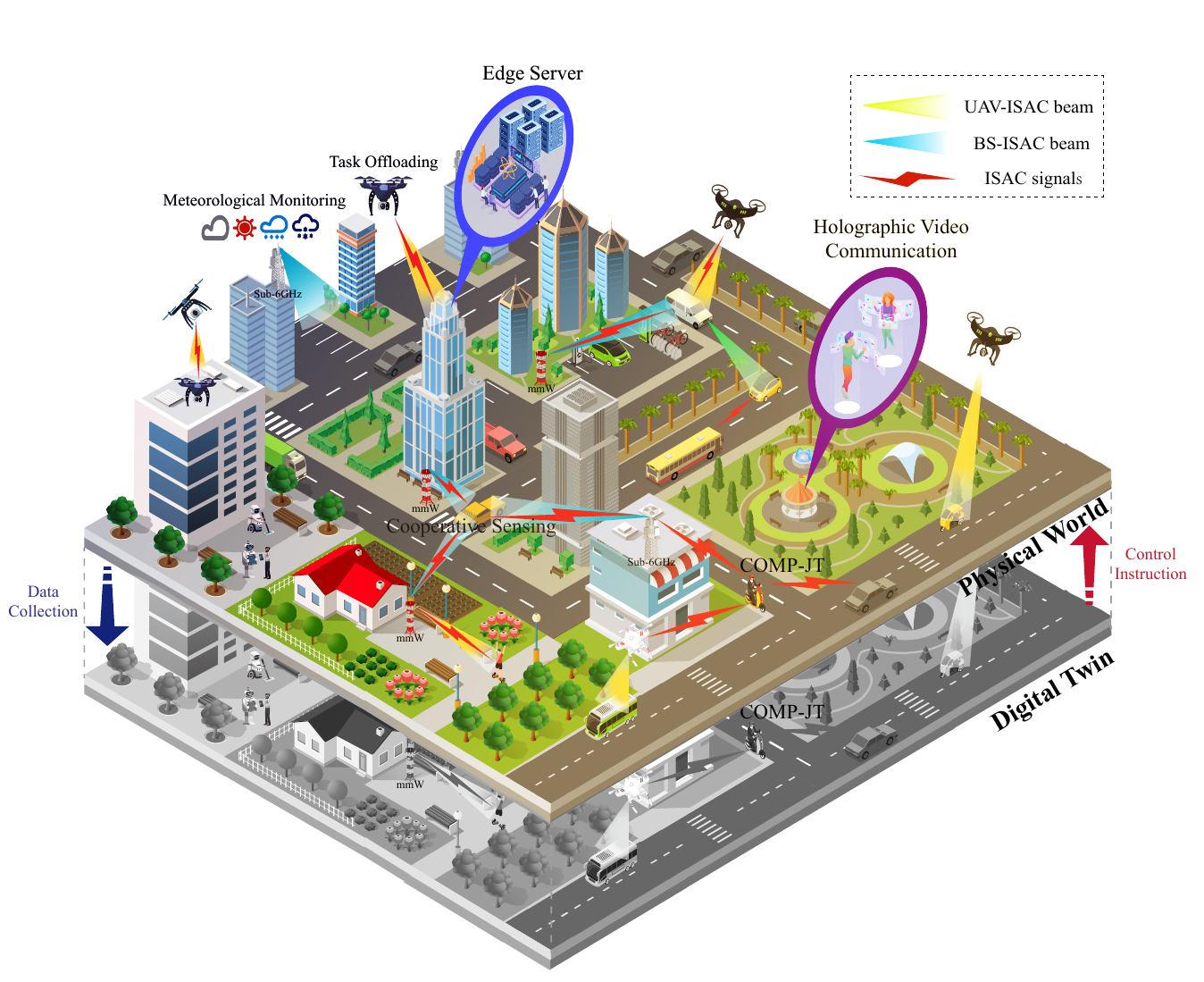}
\caption{Application Scenarios of ISAC-Assisted HDT.}
\label{FIGURE_1}
\end{figure}

\begin{table*}[!b] % 使用 figure* 环境以跨越两栏
    \begin{center}
    \centering
    \caption{Comparison of ISAC-HDT and Traditional Communication Scenarios}
    \label{fig:comparison}
    \resizebox{\textwidth}{!}{%
    \begin{tabular}{|c|c|c|c|c|}
        \hline
        \textbf{HDT Scenarios} & \textbf{Features} & \textbf{KPI} & \textbf{ISAC-HDT} & \textbf{Traditional Communication Scenario} \\ \hline
        \multirow{6}{*}{\shortstack[c]{High-immersion \\ virtual-real \\ integration}} & \multirow{3}{*}{High network quality} & Latency & $<$0.5ms & 1ms \\ \cline{3-5}
        & & Reliability & Very high & High \\ \cline{3-5}
        & & Bandwidth & For large data & For medium data \\ \cline{2-5}
        & \multirow{2}{*}{Precise 3D imaging} & Resolution & 16K & 4K \\ \cline{3-5}
        & & Accuracy & $<$1mm & $<$1m \\ \cline{2-5}
        & Strong immersion & Sensory degree & 3D immersive & 2D visual \\ \hline
        \multirow{4}{*}{\shortstack[c]{Large-capacity \\ twin networks}} & \multirow{2}{*}{Large scene data} & Bandwidth & 1Gbit/s\textasciitilde1Tbit/s & 0.1\textasciitilde1Gbit/s \\ \cline{3-5}
        & & Connection density & $>$10/m\textsuperscript{2} & $>$\(1 \times 10^{6}\)/km\textsuperscript{2} \\ \cline{2-5}
        & High intelligence & Cloud/edge AI capability & 36.8x10\textsuperscript{15} ops & 2x10\textsuperscript{15} ops \\ \hline
        \multirow{3}{*}{\shortstack[c]{Multimodal \\ embodied \\ interaction}} & \multirow{2}{*}{Multimodal interaction} & Interaction modes & Multimodal & Single mode \\ \cline{3-5}
        & & Accuracy & $<$0.01m & $<$1m \\ \cline{2-5}
        & Large interaction scale & Bandwidth & 1Gbit/s\textasciitilde1Tbit/s & Download 500MB/s, Upload 70MB/s \\ \hline
    \end{tabular}
    }
\end{center}
\end{table*}

\textcolor{black}{HDT relies on reconstructing the environment, yet fully covering cities with cameras is infeasible. In 5G and prior, positioning was the primary sensing offered by mobile communication systems. With 6G, sensing beyond positioning will be integrated, leveraging nearly ubiquitous wireless coverage. ISAC's capabilities in positioning, detection, and imaging are crucial for HDT. 5G BS and Wi-Fi, operating below 6 GHz, offer limited positioning accuracy (several decimeters to meters) and angular resolution (5° to 10°) due to bandwidth constraints. In contrast, 6G will increase millimeter-wave BS and introduce terahertz access points, millimeter-wave sensors can achieve GHz-level bandwidth, enabling centimeter-level distance resolution. For terahertz frequencies, this parameter can reach 1 to 3 millimeters \cite{5}, facilitating high-precision imaging and significantly enhancing the accuracy of HDT environmental data. Moreover, 6G aims for up to 1 Kbps/Hz spectral efficiency and peak rates of 1 Tbps, with latency reduced to 0.5 ms, enabling ultra-low latency holographic interactions. Enhanced sensing allows for improved beamforming strategies, boosting communication performance.}

\textcolor{black}{Table 1 delineates the principal application scenarios of HDT, showcasing its quintessential attributes: authentic replication, integration of virtual and real, embodied interaction, and extension of space-time \cite{11}. Leveraging ISAC and holographic display terminals, these scenarios can achieve high resolution and immersive experiences. In essence, the KPIs of ISAC-HDT manifest across three dimensions: sensing, transmission, and display. Our focal point herein is on the sensing dimension, where from an information-theoretical viewpoint, sensing capacity is of interest, such as Mutual Information (MI) and the KPIs formulated from Shannon's theorem \cite{9}; from an estimation theory perspective, attention is given to sensing probability, encompassing the Cramér-Rao Bound (CRB), Mean Squared Error (MSE), or detection probability \cite{10,13}. The precision of sensing is bandwidth-dependent, with the expansive bandwidth of 6G millimeter waves catering to the high sensing accuracy demands of the scenarios presented in Table 1.}

\subsection{Framework of ISAC-Assisted HDT}
In response to the previously outlined application scenarios, this study introduces a structured four-layer framework, as depicted in Fig. 2, designed to facilitate ISAC-assisted HDT. Organized hierarchically from the bottom to the top, the framework comprises the physical layer, the DT layer, the intelligence layer, and the holographic interaction layer.

\subsubsection{Physical Layer}
The physical layer forms the cornerstone of the ISAC-HDT system, integrating sensing and communication to share hardware and spectrum resources. The integration facilitates environmental data acquisition and transmission, streamlining system complexity and cost. \textcolor{black}{Leveraging the unique strengths of sub-6GHz, millimeter wave, and terahertz bands meets diverse KPIs in different scenarios.} Utilizing BS, UAV, and smart devices through ISAC, alongside supplementary technologies like LiDAR and camera arrays \cite{3}, the physical layer captures real-world physical information for DT construction, providing raw data for the construction of DT, thereby supporting various forms of immersive displays, including virtual reality, augmented reality, and holographic video \cite{7}. The accuracy and depth of DT are enhanced by multi-node cooperative ISAC technology, overcoming obstacles that single-point sensing faces in creating 3D images, a topic further explored in Section \Rmnum{3}.

\begin{figure}[!t]
\centering
\includegraphics[width=1\linewidth]{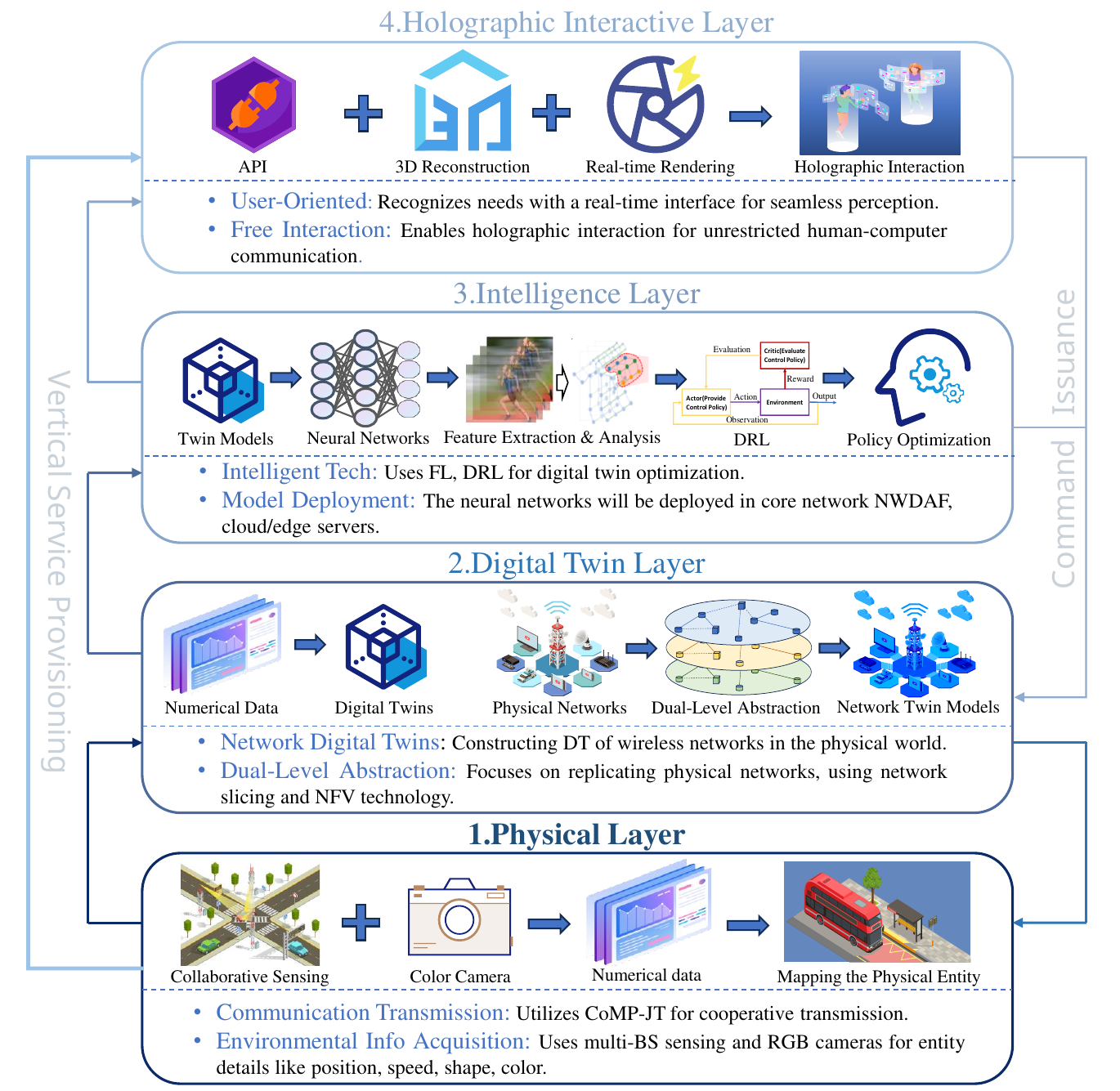}
\caption{The Four-Layer Architecture of ISAC-Assisted HDT.}
\label{FIGURE_2}
\end{figure}

\subsubsection{Digital Twin Layer}
The digital twin layer builds accurate DT models based on data collected from the physical layer, enabling digital abstraction and simulation of the physical world to provide a data foundation for upper-layer intelligent processing and analysis. As shown in Fig. 1, DT not only construct physical models of the real world but also include the ISAC networks as part of the DT, enhancing the efficient use of communication, sensing, and computing resources. DT models also support dynamic configuration and management of network slicing \cite{14}. For network operators, future 6G networks will leverage DT to build automated network operation systems. Properly designed DT models enable real-time monitoring and control of network infrastructure, facilitating analysis and optimization of the physical network, thereby improving network performance as a foundation for HDT implementation \cite{1}.

The DT layer involves two levels of abstraction: the first focuses on aggregating data from multiple nodes, updating historical data, and creating DT for new or additional users. This level balances quality of service (QoS) and quality of experience (QoE), with network slicing ensuring QoS for holographic communication services over shared infrastructure \cite{6}. The second level of abstraction aggregates data to generate network slices tailored to specific business needs, considering service requirements and resource configurations. The first level characterizes individual users and their needs, while the second level addresses network resource utilization and service capabilities, with model parameters managed by a virtual control unit.

\subsubsection{Intelligence Layer}
The intelligence layer employs neural networks to analyze DT data, extracting insights to support decision-making and predictive simulations. It incorporates deep reinforcement learning to enhance the DT model's autonomy, adaptability, and self-optimization. This layer also incorporates learning-based super-resolution techniques to enhance the resolution of environmental modeling \cite{4}, semantic communication techniques to compress data for transmission, or knowledge graph technology to better understand the interactions between various factors, supporting more precise and efficient decision-making in the ISAC-HDT system.

\begin{figure}[!t]
\centering
\includegraphics[width=1\linewidth]{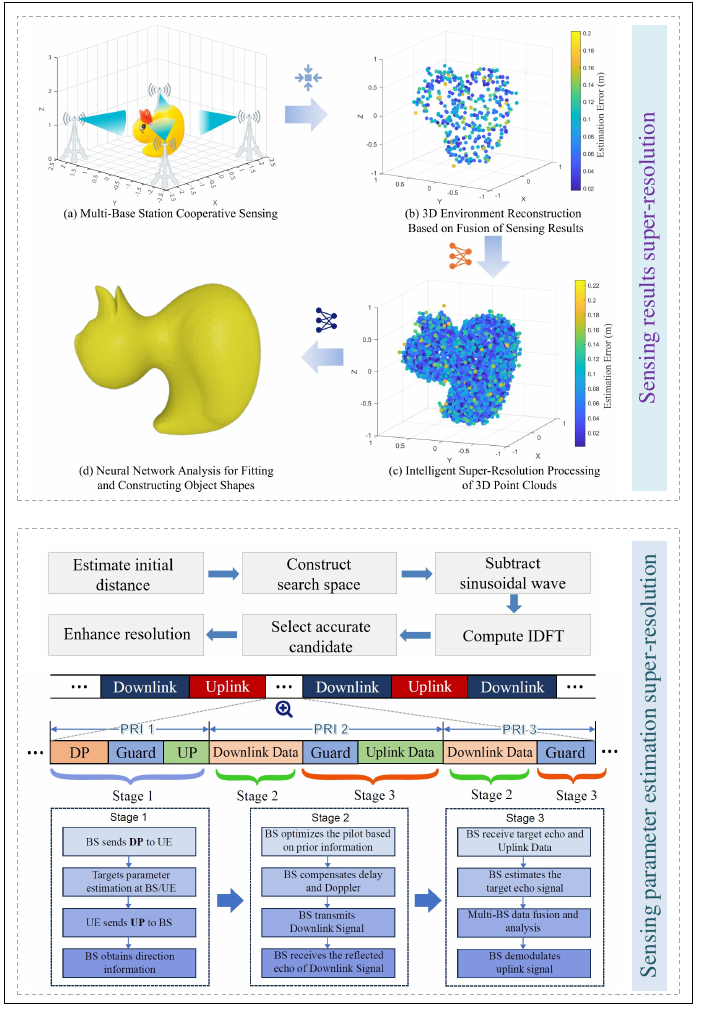}
\caption{Super-resolution for HDT construction assisted by ISAC.}
\label{FIGURE_3}
\end{figure}

\textcolor{black}{As data sensing, transmission, and analysis expand, the need to securely aggregate ISAC data from various points—often encompassing user privacy—poses severe challenges to the security of HDT. This scenario renders centralized learning obsolete, necessitating a shift towards distributed learning paradigms, with federated learning (FL) standing out as a preeminent solution \cite{14}. Initially, a global model, crafted by cloud servers, distributes its parameters to edge and local models for refinement and updates. These enhancements, once securely encrypted, ascend through layers for amalgamation, enabling iterative refinement via interaction between local and global models. This process culminates in a model adept at fusing multi-node sensing data, facilitating communication, analysis, and physical world predictions with improved efficacy.}

\subsubsection{Holographic Interaction Layer}
The results of the intelligence layer processing are intuitively displayed in the form of 3D holographic videos, providing users with an immersive interaction experience, and realizing the freedom and efficiency of human-computer interaction. The holographic interaction layer includes all application interfaces using the physical network. These applications can request services from the wireless network system based on DT and pass them to the digital twin layer. This layer also involves rendering technology. The rendering methods involved in HTC include multi-view stereo rendering technology, ultra-multi-view virtual stereo content rendering technology, and multi-plane image technology, etc. Because the holographic interaction layer directly faces the user, user needs should be clarified at this layer, and abstracted into multi-dimensional indicators to be passed to the lower layer.

\subsection{Enhancement of HDT Resolution}

Constructing HDTs with centimeter-level or higher resolution requires bandwidth beyond the capabilities of existing 5G networks, while Terahertz BS face significant challenges in output power, noise figure, and frequency stability \cite{5}. Therefore, super-resolution techniques are essential to enhance the accuracy of HDT models and improve holographic display performance. As shown in Fig. 3, super-resolution in ISAC-assisted HDT construction focuses on two aspects: parameter estimation and environment modeling, corresponding to the physical layer and intelligence layer in Fig. 2. 

The bottom of Fig. 3 illustrates a three-stage protocol and corresponding frame structures for HDT construction. Achieving super-resolution in sensing parameters relies on fully exploiting signal characteristics and adopting appropriate signal processing techniques \cite{15}, such as converting echo signals into generalized array signal forms and using improved estimation of signal parameters via rotational invariance techniques or multiple signal classification algorithms. In distance estimation, Stage 1 involves initial sensing at the BS or UE to estimate the target distance and construct a search space. In Stage 2, the BS removes echo signal components for each candidate distance and applies phase compensation to eliminate interference. Finally, Stage 3 uses inverse discrete Fourier transform to refine the distance estimate, selecting the candidate with the smallest deviation from the actual echo signal.

\begin{figure*}
  \centering
  \includegraphics[width=\textwidth]{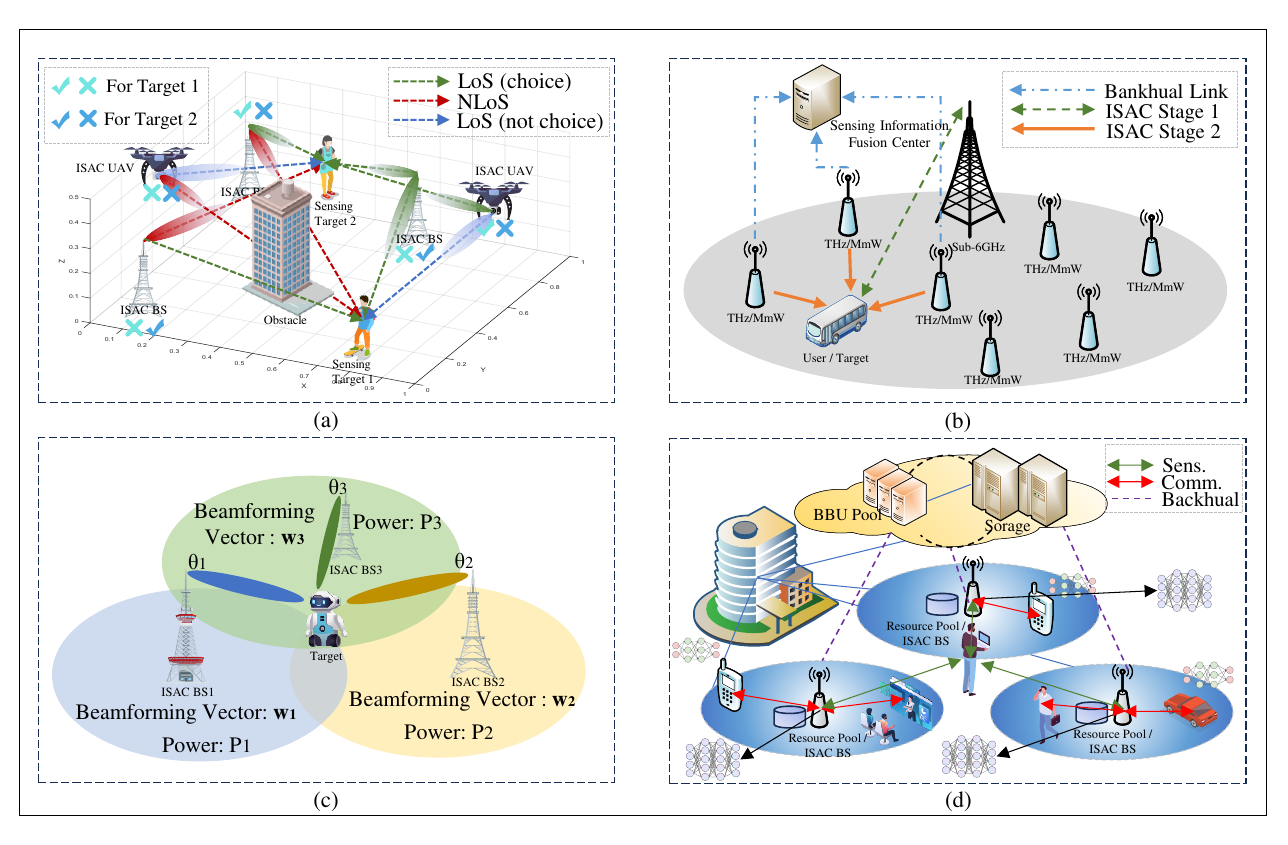}
  \caption{Perspectives of cooperative sensing technology for constructing HDT.}
  \label{FIGURE_4}
\end{figure*}

After parameter estimation, point cloud super-resolution techniques can further enhance the resolution of constructed images \cite{4}. These techniques use low-resolution images or sequences as inputs, combined with prior knowledge, to reconstruct high-resolution images. While interpolation methods can enhance resolution, they fail to recover lost details. Neural networks such as CNN and GCN can capture the topological structure and local geometric features of point clouds to generate fine-grained details. Additionally, methods based on GAN, diffusion model, and Transformer are becoming increasingly mature.

\section{Cooperative sensing for constructing HDT}
Multi-node cooperative sensing is crucial for reconstructing 3D spaces and overcoming occlusions, offering cooperative space gain and joint processing advantages in S\&C. Utilizing ISAC for HDT construction necessitates cooperative ISAC, which enables high-precision, super-resolution sensing by coordinating multiple nodes and integrating multidimensional information. This collaboration allows for the reception of reflected signals from various angles and distances, enhancing target positioning accuracy through the reduction of random errors and the utilization of multidimensional measurements. \textcolor{black}{In \cite{9}, compared to single BS sensing, the accuracy of distance and speed estimation improved by 40\% and 72\%.}

In multi-node collaboration, synchronization is critical for sensing performance \cite{13}. Nanosecond-level errors can cause meter-level distance estimation inaccuracies, as sensing requires higher precision than communication. While multi-node sensing works with current network architectures, the existing systems designed for communication fall short for precise localization. Frequency and phase differences among BS clocks, propagation delays, and hardware inconsistencies cause time reference mismatches and synchronization offsets. Methods like using the time difference of arrival between reference paths and echoes can mitigate these errors.

\subsection{ISAC Nodes Selection}
The selection of sensing nodes is vital for efficient 3D environmental information collection by ISAC, aiming to involve the minimum number of nodes while maximizing cooperative gain, as illustrated in Fig. 4(a). Key considerations for node selection include:

Selecting sensing nodes requires considering the following aspects:

\begin{itemize}
\item{Node Coverage: Consider propagation loss, multipath effects, and obstacles for NLoS impacts in urban HDT 3D mapping.}
\item{BS Configuration: Higher frequencies improve data rate and resolution but limit range and signal quality. Consider BS's hardware, including sensitivity, speed, capacity, and antenna design.}
\item{System Complexity: Balance ISAC coordination, positioning algorithms, and data fusion complexity with S\&C performance when selecting BS.}
\end{itemize}

Designing a multi-node cooperative ISAC system typically aims to optimize energy efficiency, communication, or sensing performance, or a combination thereof. Sensing performance can be evaluated through various metrics, such as the Cramér-Rao Lower Bound (CRLB), sensing MI, and radar estimation information rate. Reference \cite{8} presents an ISAC node selection strategy that minimizes total transmission power while satisfying CRLB and communication rate requirements, offering a strategic approach for BS selection in the ISAC-HDT system. However, a more comprehensive approach to node selection, considering additional factors and the management of node mobility in multi-node ISAC networks, is necessary for further advancement.

\subsection{Multi-frequency Collaboration}
Next-generation mobile networks will integrate high-frequency and low-frequency BSs for ISAC multiband collaboration \cite{1}. Low-frequency BSs provide wide coverage for initial sensing and anomaly detection, while high-frequency BSs offer precise sensing and high-speed communication without interruption, as shown in Fig. 4(b). Low-frequency BSs identify approximate target locations and user needs, supporting high-frequency BSs in refining services, with data fusion performed on servers. This collaboration combines the propagation benefits of different frequencies, mitigates interference and fading, and leverages subcarrier aperture gains from multiband channel state information (CSI) measurements for enhanced performance.

\subsection{Cooperative Beamforming}
In multi-node collaborative ISAC systems, cooperative beamforming is a crucial technique for achieving high transmission rates and precise target estimation \cite{13}, as shown in Fig. 4(c). Spatial synchronization is a key aspect of cooperative beamforming and a prerequisite for fusing sensing information. The primary challenges of spatial synchronization lie in unifying the coordinate systems of sensing information from different nodes and ensuring alignment between the sensing regions detected by these nodes. By formulating beamforming optimization problems that account for time synchronization errors, the negative impact on sensing accuracy can be mitigated. In millimeter-wave and terahertz frequency bands, the large signal bandwidth causes the signal wavelength to vary with the subcarrier index, a phenomenon known as beam squint. This results in phase differences across different antenna array elements, affecting synchronization performance. Phase compensation and similar methods are required to address these challenges.

\subsection{Multi-node Data Fusion}
Data fusion, pivotal in enhancing the benefits of multi-node cooperation, can be categorized into three types:

\begin{itemize}
    \item \textbf{Front-end fusion:} Front-end fusion involves combining raw channel information to improve signal-to-noise ratio (SNR) by coherently fusing echo signals from multiple BSs. While it ensures maximum accuracy by avoiding additional signal processing errors, it requires significant server resources and introduces high delays, making it unsuitable for delay-sensitive HDT systems. This method demands high communication capacity and precise spatiotemporal synchronization, including unified coordinate systems and time-aligned sampling to reduce clock offset errors.
    \item \textbf{Mid-end fusion:} Mid-end fusion focuses on channel feature quantities, where receiving nodes extract features and send them to the server, reducing data feedback and achieving moderate accuracy. It also requires unified spatial coordinate systems and synchronized sampling times across multiple BSs.
    \item \textbf{Back-end fusion:} Back-end fusion processes sensing results locally at receiving nodes, with the server performing clustering and weighted averaging. However, errors uncorrected at earlier stages can accumulate, reducing the accuracy of digital twins and impacting holographic interactions.
\end{itemize}

Mid-end fusion emerges as the most appropriate choice for constructing ISAC-HDT systems, contingent upon the synchronization of the receiving node's position and time. Achieving such synchronization, particularly with mobile communication BSs as receiving nodes, presents significant challenges. To overcome this, a symbol-level cooperative sensing method, suitable for the synchronization capabilities of mobile communication systems, was proposed in \cite{9}. This method involves preprocessing the signal from a single BS, phase adjustment and accumulation of symbols, calibration of each BS's symbol vector phase, and finally, symbol-level multi-node sensing information fusion through compensation and multiplication of the symbol vector to enhance fusion quality.

As the volume of sensed and analyzed data grows, the importance of privacy protection escalates, making centralized data learning and fusion untenable. FL offers a solution by enabling the training of machine learning models without the need to share original data, as depicted in Fig. 4(d). A framework introduced in \cite{14} utilizes Internet of Things (IoT) devices for physical world data collection, employing FL for multi-source data fusion and DT model training. This approach, equipping user smart devices and BSs with neural networks for model training capability, represents an AI-empowered mid-end fusion. Upon training completion, model updates are uploaded to the server for aggregation and distribution using FedSGD or FedAvg, enhancing the global model's generalization ability and suitability for multi-node cooperation with heterogeneous devices.

\section{Simulation and Discussion}

\begin{figure}[!h]
  \centering
  \includegraphics[width=1\linewidth]{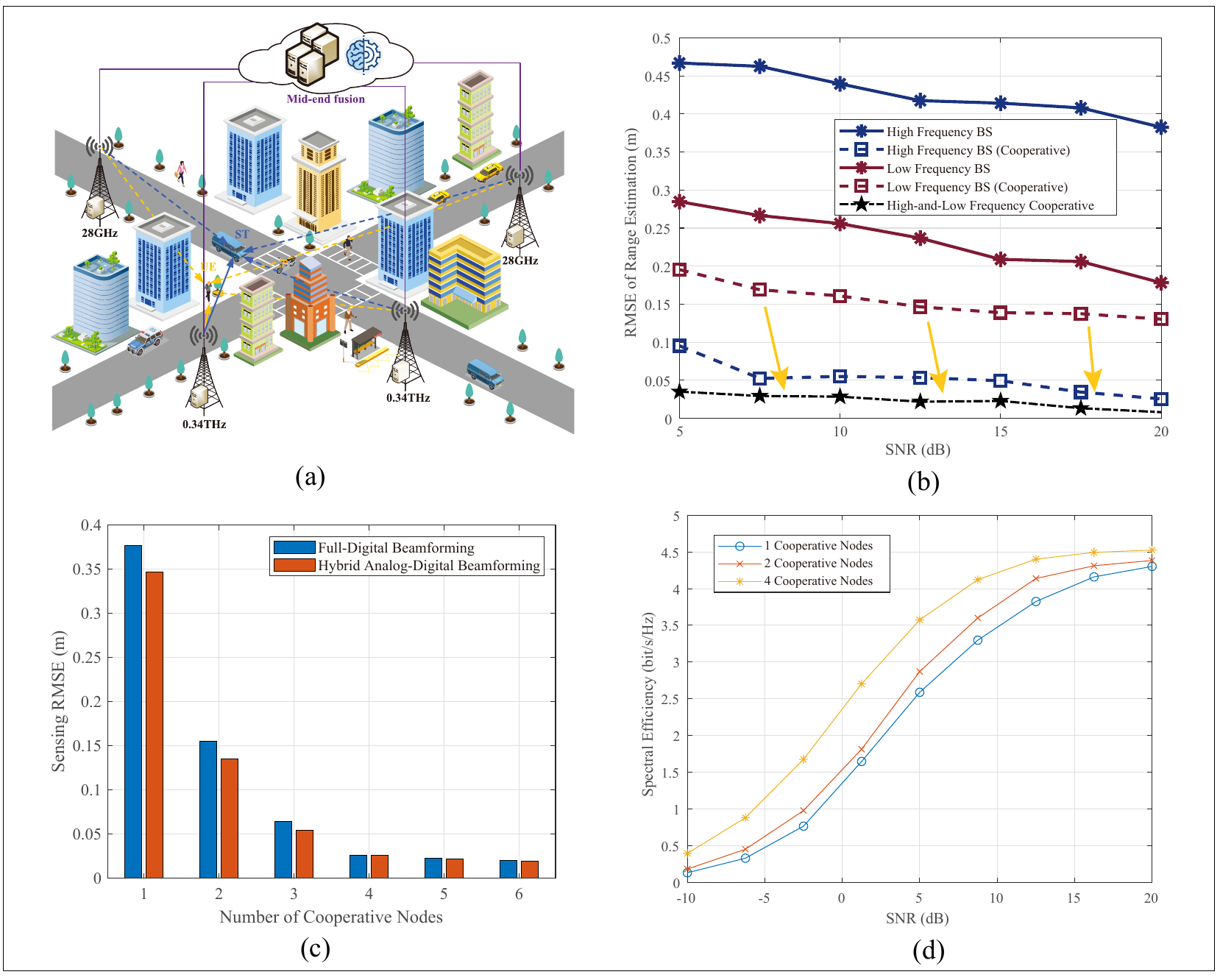}
  \caption{Performance improvement of S\&C in multi-node collaboration.}
  \label{FIGURE_5}
\end{figure}

\subsection{Multi-Frequency Cooperative Gain}
In the scenario depicted in Fig. 5(a), there are two high-frequency BSs operating at a carrier frequency of 0.34 THz with a bandwidth of 1 GHz, alongside two low-frequency BSs operating at a carrier frequency of 28 GHz with a bandwidth of 95 MHz. The urban landscape features randomly distributed buildings of various sizes and locations, which obstruct ISAC signals, leading to NLoS paths and consequently, a higher path loss exponent. A mid-End fusion approach serves for the integration of sensing information. The results from 500 Monte Carlo simulations, as shown in Fig. 5(b), indicate that relying solely on high-frequency BSs without cooperation for distance estimation tends to result in a higher probability of NLoS occurrences due to obstructions, negatively impacting both received power and sensing accuracy. Conversely, cooperation among high-frequency BSs significantly increases the likelihood of establishing LoS channels, thereby markedly improving sensing accuracy. However, the resolution enhancement potential of single BS or cooperative low-frequency BSs remains limited due to bandwidth constraints. Furthermore, the smaller difference in loss between NLoS and LoS paths means that the benefits of cooperation are less pronounced. Incorporating some low-frequency BSs into the cooperative framework with high-frequency BSs for sensing tasks offers a more robust solution, ensuring stable sensing performance from low-frequency BSs that are less susceptible to obstructions, even in scenarios where all high-frequency BSs are blocked.

\subsection{Multi-node Cooperative Gain}

In this subsection, we discuss the advantages of cooperation among different BSs in the same obstacle-rich environment. The analysis focuses on a scenario involving a single user, with all four BSs operating at 0.34 THz and utilizing OFDM waveforms. As illustrated in Fig. 5(c), the distance estimation root mean square error (RMSE) transitions from sub-meter to millimeter accuracy as the number of sensing BS increases. However, the marginal benefits decrease with the addition of more BSs, primarily because the initial increase in BSs compensates for the lack of spatial diversity, but further additions lead to more complex and severe beam interference. Fully digital beamforming, with its precise and flexible control over phase and amplitude, outperforms hybrid beamforming, although the difference is slight. The latter maintains performance while reducing the number of RF chains and power consumption. Perfect CSI estimation is assumed for signal reception, and a phase pre-compensation factor is introduced to facilitate coherent joint transmission, addressing the stringent synchronization requirements for carrier frequency and phase among different transmitters. Fig. 5(d) highlights the spectral efficiency improvements achieved through the power and spatial diversity gains provided by coherent joint transmission using CoMP technology.

\subsection{Power Control in Multi-node Cooperation}

Simultaneously, Research on the energy consumption of cooperative BSs has been gaining increasing attention. In \cite{12}, by fine-tuning the beam scheduling among various cooperative BSs, the energy efficiency benefits of each station are fully exploited, thus minimizing the overall energy consumption of the network. In the simulation presented in Fig. 6, we explore the development of cooperative beamforming strategies for multi-node collaborative communication and sensing, subject to SNR and the CRLB constraints for sensing accuracy. The goal is to achieve the stipulated requirements with the minimum transmission power by optimizing the hybrid beamforming variables of BSs. The simulation setup includes a bandwidth of 1 MHz, a carrier frequency of 6 GHz, adopts a Rician channel model with a K-factor of 5 dB, a CRLB constraint lower bound of 0.03 $\text{m}^2$, and a noise power spectral density according to the standard benchmark of $-$174 dBm/Hz. We compare the CRLB approximation method with the Semidefinite Relaxation (SDR) method. While the SDR method transforms the optimization problem into a convex one, increasing computational complexity, the CRLB approximation method simplifies the sensing constraint to a linear form, thereby reducing the problem's complexity. Simulation results indicate that under stringent signal-to-interference-plus-noise ratio (SINR) conditions, the CRLB approximation method performs effectively, and more demanding performance criteria, such as a reduced CRLB, necessitate greater transmission power. This highlights the importance for practical ISAC-HDT systems to carefully balance the design considerations among transmission power, the number of nodes, and performance requirements.

\begin{figure}[!t]
  \centering
  \includegraphics[width=1\linewidth]{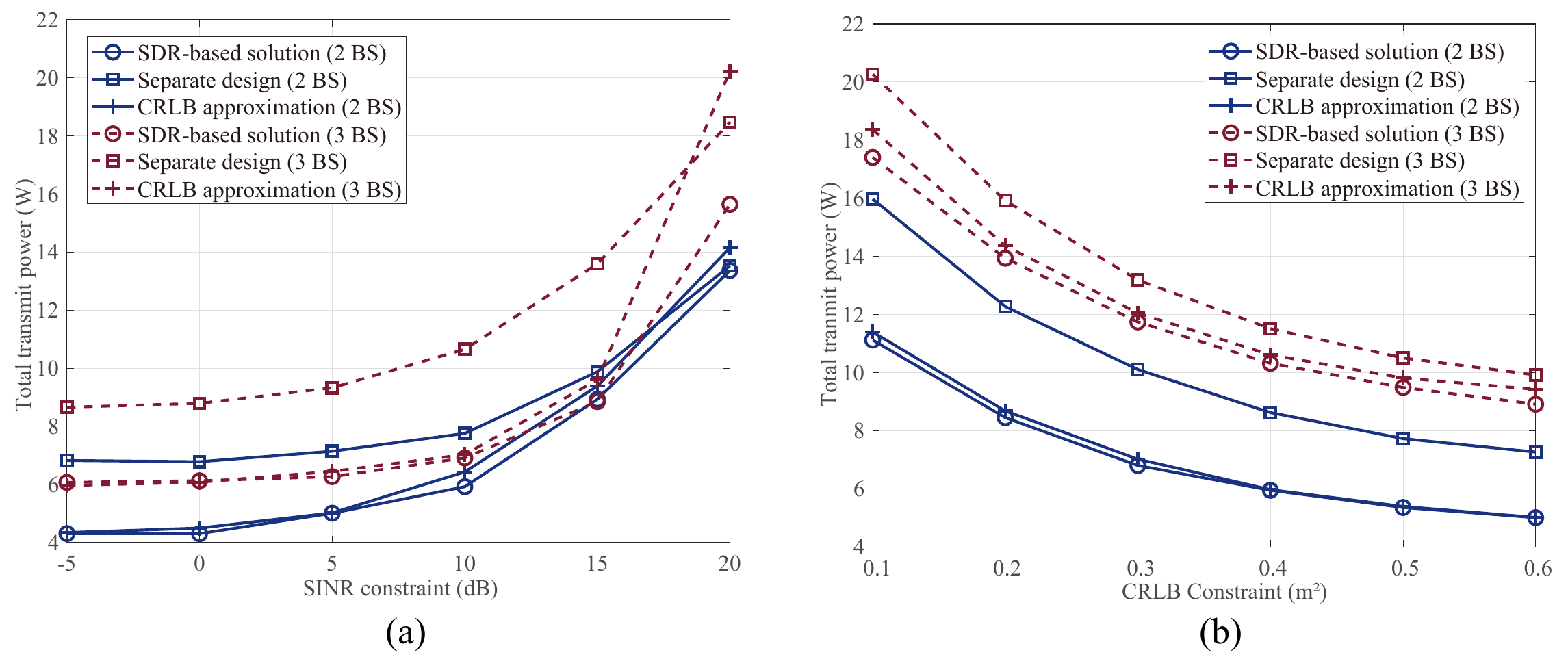}
  \caption{Minimum transmission power under different S\&C performance constraints for varying numbers of cooperative BSs.}
  \label{FIGURE_6}
\end{figure}

\section{Conclusions and Future Directions}
In this paper, we introduced the application scenarios of ISAC-assisted HDT and proposed a structured four-layer ISAC-HDT architecture tailored to these scenarios, elucidating the technology of each layer to underscore its benefits. In addressing the issue where the resolution of ISAC falls short of meeting the requirements for HDT, we explore the introduction of super-resolution techniques at both the physical and intelligence layers to enhance the resolution of parameter estimation and point cloud construction. Next, our focus shifts to multi-node cooperative sensing technology for acquiring physical layer sensing information, from which we derive four perspectives to aid in the precise construction of HDT, thereby offering robust technical support for harnessing ISAC in sourcing data for HDT. We conclude by identifying several cutting-edge technologies poised to advance the ISAC-HDT system, marking them as avenues for future research.

\subsection{Ubiquitous Computing and Ambient Intelligence}
\textcolor{black}{Ubiquitous computing, aimed at fully sensing and computing the environment, benefits greatly from ISAC technology by providing a seamless, real-time support platform for ISAC-HDT systems. It extends to ambient intelligence by adding human-centric computational interactions, focusing on smart decision-making and automatic responses. Ambient intelligence enhances the environment through the intent-driven integration of multimodal sensory data, offering a personalized, human-centric interaction experience.}

\subsection{Over-the-Air Computation}
\textcolor{black}{During deep learning model training in the intelligence layer, FL is employed for iterative updates of both local and global models. However, the time consumed in the upload and aggregation process of updated local models can significantly impact the real-time holographic interaction experience for users. Over-the-air computation enables real-time aggregation of multi-node sensory data in ISAC-HDT systems, leveraging the waveform superposition property of multiple access channels to perform function computation in the air, directly obtaining the results at the receiver. Efficiently aggregating local training outcomes of distributed sensor data through AirComp can enhance FL performance, contributing to the efficiency of multi-point cooperative sensory data fusion and global model aggregation and updates.}

\subsection{Large Language Models and Semantic Communication}
\textcolor{black}{Multimodal cooperative sensing, particularly with video sensor support, leads to the transmission of vast sensory data, increasing node dprocessing burdens and latency challenges. Semantic communication significantly reduces the data transmission volume \cite{2}. Yet, existing schemes face issues like semantic modeling and optimization difficulties, limited scheme knowledge, and constrained semantic understanding. The advent of large language models presents a new solution, utilizing their robust semantic comprehension and generation to facilitate easier data compression, transmission, and recovery for DT, acting as an advanced method for data feature extraction and abstraction, thereby boosting data transmission efficiency.}

% \section*{Acknowledgments}
% This work is supported in part by the National Natural Science Foundation of China under Grants 62225103, U22B2003, and U2441227; in part by the Beijing Natural Science Foundation under Grant L241008; in part by the Fundamental Research Funds for the Central Universities under Grant FRF-TP-22-002C2; in part by the National Key Laboratory of Wireless Communications Foundation under Grant IFN20230201, and in part by the Xiaomi Fund of Young Scholar. 


\bibliographystyle{IEEEtran}

\section{Biography}
\vspace{-10 mm}
\begin{IEEEbiographynophoto}{Haijun Zhang}
 is currently a Full Professor with the University of 
Science and Technology Beijing, China. He was a postdoctoral 
research fellow with the Department of Electrical and Computer 
Engineering, University of British Columbia (UBC), Canada. He 
serves as an editor for IEEE Trans. Information Forensics and 
Security and IEEE Trans. Network and Service Engineering.
\end{IEEEbiographynophoto}
\vspace{-10 mm}
\begin{IEEEbiographynophoto}{ZiYang Zhang}
received the B.S. degree from the School of Computer and Communication Engineering, University of Science and Technology of Beijing, Beijing, China, in 2023. He is currently pursuing the M.S. degree at University of Science and Technology Beijing, China. His research interests include immersive communication and integrated communication and sensing.
\end{IEEEbiographynophoto}
\vspace{-10 mm}
\begin{IEEEbiographynophoto}{Xiangnan Liu}
received the B.S. degree and the Ph.D. degree from the School of Computer and Communication Engineering, University of Science and Technology of Beijing, Beijing, China, in 2019 and 2024, respectively. He is a Postdoctoral Research Fellow in Department of Electrical and Computer Engineering, KTH Royal Institute of Technology, Stockholm, Sweden. His research interests include access control, beamforming, and resource allocation in next generation wireless communication.
\end{IEEEbiographynophoto}
\vspace{-10 mm}
\begin{IEEEbiographynophoto}{Wei Li}
is currently a Full Professor in University of Science and Technology Beijing, China. Her research interests include mobile communication, industrial internet, intelligent manufacturing.
\end{IEEEbiographynophoto}
\vspace{-10 mm}
\begin{IEEEbiographynophoto}{Haojin Li} received the M.S. degree in information and communication engineering from the University of China Academy of Telecommunication Technology, China, in 2020. He joined Sony China Laboratory in 2020 and is responsible for 5G System Level Simulator development, in June 2022, he was promoted to Deputy Principal Research and Development Researcher. He is currently working toward the Ph.D. degree in Electronics Information with the School of Computer and Communication Engineering, University of Science and Technology Beijing, China. He is currently a Deputy Director of Sony and USTB Joint Laboratory. His research interests include integrated sensing and communication.
\end{IEEEbiographynophoto}
\vspace{-10 mm}
\begin{IEEEbiographynophoto}{Chen Sun}
received the Ph.D. degree in electrical engineering from Nanyang Technological University, Singapore, in 2005. He contributed to IEEE standards and Wi-Fi Alliance specs. Now he's the deputy head of Beijing Lab at Sony R\&D Center. His research interests include smart antennas, cognitive radio, V2X, federated learning, and AI in wireless networks. He received the
 IEEE Standards Association IEEE 1900.6 Working Group Chair Award for
 leadership in 2011, the IEEE 802.11af Outstanding Contributions Award in
 2014, and the IEEE 802.19.1 Outstanding Contributions Award in 2018.
\end{IEEEbiographynophoto}
\vfill

% \clearpage
% \begin{figure*}
% \centering
% \includegraphics[width=1\linewidth]{HDT/pic1.pdf}
% \caption{Application Scenarios of ISAC-Assisted HDT.}
% \label{FIGURE_1}
% \end{figure*}

% \clearpage
% \begin{figure*}
% \centering
% \includegraphics[width=1\linewidth]{HDT/pic2.pdf}
% \caption{The Four-Layer Architecture of ISAC-Assisted HDT.}
% \label{FIGURE_2}
% \end{figure*}

% \clearpage
% \begin{figure*}
% \centering
% \includegraphics[width=0.75\linewidth]{HDT/pic3.pdf}
% \caption{Super-resolution for HDT construction assisted by ISAC.}
% \label{FIGURE_3}
% \end{figure*}

% \clearpage
% \begin{figure*}
% \centering
% \includegraphics[width=\textwidth]{HDT/pic4.pdf}
% \caption{Perspectives of cooperative sensing technology for constructing HDT.}
% \label{FIGURE_4}
% \end{figure*}

% \clearpage
% \begin{figure*}
% \centering
% \includegraphics[width=1\linewidth]{HDT/pic5.pdf}
% \caption{Performance improvement of S\&C in multi-node collaboration.}
% \label{FIGURE_5}
% \end{figure*}

% \clearpage
% \begin{figure*}
% \centering
% \includegraphics[width=1\linewidth]{HDT/pic6.pdf}
% \caption{Minimum transmission power under different S\&C performance constraints for varying numbers of cooperative BSs.}
% \label{FIGURE_6}
% \end{figure*}

\end{document}